\documentstyle[epsf,prd,aps,floats]{revtex}

\input{psfig.tex}

\begin{document}

\title{A Cosmological Tale of Two Varying Constants}
\author{John D. Barrow$^1$, Jo\~ao Magueijo$^2$ and H\aa vard B. Sandvik$^2$}
\address{$^1$DAMTP, Centre for Mathematical Sciences,\\
Cambridge University, Wilberforce Rd.,\\
Cambridge CB3 0WA, UK.\\
$^2$Blackett Laboratory, Imperial College,\\
Prince Consort Rd., London SW7 2BZ, UK}
\date{The Date}
\maketitle

\begin{abstract}
We formulate a simple extension of general relativity which
incorporates space-time variations in the Newtonian gravitation
'constant', $G$, and the fine structure 'constant', $\alpha $,
which generalises Brans-Dicke theory and our theory of varying
$\alpha .$ We determine the behaviour of Friedmann universes in
this theory. In the radiation and dust-dominated eras $\alpha G$
approaches a constant value and the rate of variation of $\alpha $
is equal to the magnitude of the rate of variation in $G$. The
expansion dynamics of the universe are dominated by the variation
of $G$ but the variation of $G$ has significant effects upon the
time variation of $\alpha .$ Time variations in $\alpha $ are
extinguished by the domination of the expansion by spatial
curvature or quintessence fields, as in the case with no $G$
variation.
\end{abstract}

\section{Introduction}

There have been many studies of the cosmological consequences of allowing
some of the traditional constants of Nature to change. These include
evaluations of the effects of altering the observed value of a constant to
another constant value and studies of the time-evolution of 'constants' in
generalisations of the general theory of relativity that allow them to
become space-time variables. The most studied case is that of varying the
Newtonian gravitation constant, $G$, through the Brans-Dicke (BD)
scalar-tensor theory of gravity \cite{bd}. Recently, following Bekenstein,
\cite{bek}, Sandvik, Barrow and Magueijo \cite{sbm},\cite{bsm} have
developed a theory (BSBM) which describes the space-time variation of the
fine structure constant. In these, and other, studies of varying constants
only a single constant is allowed to vary at one time. However, since we
have no understanding of why the constants of Nature take the values that
they do, whether they are logically independent, or even whether they all
are truly constant, this restriction is somewhat artificial. Motivated by
recent observational evidence for a time evolution of the fine structure
'constant', $\alpha ,$ at redshifts $z\sim 1-3.5$, \cite{webb99}, \cite
{webb01}, \cite{murphy}, we have unified the BD and BSBM theories to produce
an exact theory which describes the simultaneous variation of $\alpha $ and $%
G$. This type of model also provides a framework within which to consider
the consequences of changes in the scale of extra dimensions of space on
apparent three-dimensional coupling constants.

In section 2 we set up the theory and evolution equations for
Friedmann universes in a theory that generalises general
relativity to include varying $\alpha $ and $G.$ In section 3 we
show how to find the cosmological solutions during the
dust-dominated eras. We find an exact solution where $\alpha G$ is
constant during the dust era while $\alpha $ and $G^{-1}$ both
increase with time. We then determine analytically the coupled
evolution of $\alpha $ and $G$ during the radiation, curvature,
and vacuum-energy dominated eras of cosmological expansion. From
here we go on to check the solutions numerically and we show how
in Universes like our own, with actual initial values for $\alpha$
and $G$ the asymptotic behaviour is never reached. Instead we find
constant $\alpha$ and $G$ in the radiation era, slow growth of
$\alpha$ and slow decrease in $G$ in the dust epoch, constant
values for both in curvature dominated universe, and constant
$\alpha$ and decreasing $G$ in $\Lambda$ dominated epoch.
Generally we find that the overall evolution of the expansion
scale factor of the universe is dictated by the $G$ variation and
assumes the form found in the Brans-Dicke theory to a very good
approximation irrespective of the $\alpha $ variation. The
evolution of $\alpha $ is influenced by the $G$ variation but does
not differ much from that found in the BSBM cosmologies where only
$\alpha $ varies.



\section{Field Equations}

We introduce the structure of the BSBM theory for varying $\alpha
$ as another matter field in Brans-Dicke theory. The resulting
theories has two scalar fields: the BD field $\phi $ propagating
variations in $G,$ and the field $\psi $ propagating variations in
$\alpha$. The action for this theory becomes
\begin{equation}
S=\int d^4x\sqrt{-g}\left( R\phi +\frac{16\pi }{c^4}{\cal
L}-\omega _{BD} \frac{\phi _{,\mu }\phi ^{,\mu }}\phi \right)
\label{act}
\end{equation}
where
\begin{equation}
{\cal L}={\cal L}_m+{\cal L}_{em}\exp (-2\psi )+{\cal L}_\psi ,  \label{lag}
\end{equation}
and
\[
{\cal L}_\psi =-\frac \omega 2\psi _{,\mu }\psi ^{,\mu }.
\]

The field equations for the theory, specialised to the case of a
homogeneous and isotropic Friedmann space-time metric containing
dust and radiation perfect fluids are:
\begin{eqnarray}
3\frac{\dot a^2}{a^2} &=&\frac{8\pi }\phi \left( \rho _m(1+|\zeta|
\exp (-2\psi ))+\rho _r\exp (-2\psi )+\rho _\psi \right)
-3\frac{\dot a^{\ }\dot
\phi }{a\ \phi }+\frac{\omega _{BD}}2\frac{\dot \phi ^2}{\phi ^2}-\frac k{a^2%
}  \label{fried} \\
\ddot \phi +3\frac{\dot a}a\dot \phi &=&\frac{8\pi }{3+2\omega _{BD}}(\rho
_m-2\rho _\psi )  \label{phi} \\
\ddot \psi +3H\dot \psi &=&-\frac 2\omega \exp (-2\psi )\zeta \rho
_m
\label{ddotpsi} \\
\dot {\rho _m}+3H\rho _m &=&0  \label{dust} \\
\dot {\rho _r}+4H\rho _r &=&2\dot \psi \rho _r  \label{rad}
\end{eqnarray}
where $\rho _\psi =\frac \omega 2\dot \psi ^2$ is the kinetic
energy density for the $\psi $ fluid, with $\omega$ the coupling
setting the relevant energy scale for the $\psi$-field. $\zeta $
is defined as the ratio ${\cal L}_{em} / \rho_m$ averaged over all
types of matter in the universe. The fine structure 'constant' is
given by ($\hbar =c=1$)

\begin{equation}
\alpha \equiv \alpha_0\exp (2\psi ),  \label{alf}
\end{equation}
where $\alpha_0$ is the present day value of the fine structure
'constant'. The present-day value of $G$ is set equal to unity.

We shall confine our attention to the case with $\zeta <0$ where the
magnetic energy dominates the electric field energy of the matter coupling
to electric charge in the universe. This places particular constraints upon
the nature of the cold dark matter dominating the universe today. From our
earlier studies, \cite{sbm}-\cite{bsm}, we know that this case provides a
slow variation with $\alpha $ increasing logarithmically in time during the
dust era but staying constant during any subsequent curvature or
cosmological constant dominated era. Also, in a universe with a
matter-radiation balance like our own, $\alpha $ remains constant during the
radiation era except close to the initial singularity. Negative $\zeta $
models are well behaved and correspond to the dark matter in the universe
being dominated by magnetic coupling, (for example superconducting cosmic
strings contribute $\zeta =-1$). The expansion scale factor evolution is not
affected by variations in $\alpha $ to leading order. By contrast, the
choice $\zeta >0$ creates major changes to cosmological evolution. It does
not lead to slow increase of $\alpha $ with time during the dust era, as
observations suggest, and the evolution of the expansion scale factor is
affected to leading order (see for example refs.\cite{olive,zald} who
discuss related theories for the variation of $\alpha $ with $\zeta >0$ and
hence $\dot \alpha <0$ cosmological behaviour in the dust era in contrast to
our discussions in \cite{sbm}-\cite{bsm} and below). In what follows we
shall investigate how the $\zeta <0$ evolution of the fine structure
constant couples to variation of $G$ in the Brans-Dicke theory.

The constant $\omega _{BD}$ is the Brans-Dicke parameter and $\omega $ is
the analogous parameter for the coupling of the $\psi $ field driving
variations in $\alpha .$ We have used the facts that dust is pressureless, $%
p=\rho _r/3$ for a sea of radiation and $p=\rho _\psi $ for a fluid with
kinetic energy only. Equation (\ref{fried}) can be recast for numerical
solutions
\begin{equation}
\frac{\dot a}a=-\frac 12\frac{\dot \phi }\phi \pm \frac 12\sqrt{\left( \frac{%
\dot \phi }\phi \right) ^2+\frac 43\left( \frac{8\pi \rho }\phi +
\frac {\omega_{BD}} 2 \left(\frac{\dot \phi }\phi \right)
^2\right) -4\frac k{a^2}} \label{num}
\end{equation}
and eqn. (\ref{rad}) integrates to give $\rho _\gamma \exp (-2\psi
)\propto a^{-4}.$ Note that this is the combination that appears
in the generalised Friedmann equation, (\ref{fried}). In ref.
\cite{bsm} we showed how to deduce the solutions of these
equations when $G$ is constant. Here, we will extend this analysis
to the new situation where both $\alpha $ and $G$ vary in time.

\section{Dust era evolution}

From our study of the Friedmann models in BSBM theory we know
that, to a very good approximation, the $\alpha $ variations do
not significantly affect the evolution of the expansion scale
factor $a(t)$. The effects of varying $G$ in Brans-Dicke theories
is different. No matter how slow the variation in $G$, a
correction will occurs to the power of the time-variation of the
expansion scale factor. In the dust era we assume the asymptotic
solution for the Brans-Dicke (BD) flat dust model holds to high
accuracy. This is an exact solution of (\ref{fried}) for $\zeta
=k=\psi =\rho _\gamma =0$ and is the late-time attractor of the
general flat BD dust
solution (see refs. \cite{nar,fink}, \cite{bar}) which differs only as $%
t\rightarrow 0$, where the solution becomes dominated by the
kinetic energy of the $\phi $ field and approaches the BD vacuum
solution. Thus, to leading order the expansion dynamics and $\phi
$ evolution are described at late times by the exact Brans-Dicke
dust solution with $k=0$:
\begin{eqnarray}
a(t) &\propto &t^{(2-n)/3};\vspace{2cm}\phi =\phi _0t^n  \label{bd1} \\
\rho &=&Ma^{-3}\hspace{1.0in}\dot M\equiv 0  \label{bd2} \\
\phi _0 &\equiv &\frac{8\pi M}{n(3+2\omega _{BD})}  \label{bd3}
\end{eqnarray}
where $n$ is related to the Brans-Dicke parameter by
\begin{equation}
n\equiv 2/(4+3\omega _{BD}),  \label{bd4}
\end{equation}
and $M$ is the present density of the universe in Planck units, $M
\sim 10^{-123}$.

What is the asymptotic solution for $\alpha $ during the dust era?
The relevant equation is (\ref {ddotpsi}), which can now be
rewritten as
\begin{equation}
\frac d{dt}(t^{2-n}\dot{\psi})=N\exp (-2\psi )  \label{psi}
\end{equation}
where

\[
N\equiv -\frac{2\zeta }\omega \rho _ma^3=-\frac{2\zeta M}\omega >0%
\hspace{1.0in}\dot N\equiv 0,
\]
and $-\zeta / \omega \approx 10^{-4}$ is the best fit of this
parameter ratio to the observations of Webb et. al.
\cite{webb99}-\cite{murphy}\footnote{The value used for $-\zeta /
\omega$ is the value fitted for the BSBM theory with constant $G$.
However, since $n$ is so close to zero it should not be
significantly different numerically in the case with varying $G$
}.

Unlike the case with constant $G$, there is an exact solution (for
$\omega _{BD}$ positive and finite)
\begin{equation}
\psi (t)=\frac n2\ln (t)+\frac 12\ln N-\frac 12\ln (\frac n2-\frac{n^2}2)
\label{psi1}
\end{equation}
so we have, using this solution for $\psi $ to solve for $\phi $ in (\ref
{phi}):
\begin{equation}
\alpha (t)=\alpha_0 \exp (2\psi )\ =\alpha_0 \frac{2Nt^n}{n(1-n)}
\label{alf2}
\end{equation}
Hence, there is a simple relationship between $\alpha (t)$ and $G(t)$:

\begin{equation}
\phi =G^{-1}=\frac{2\pi \omega (1-n) }{-\zeta (3+2\omega _{BD})}
\frac \alpha {\alpha_0}=\frac{ 2\pi \omega (2+3\omega
_{BD})}{-\zeta (3+2\omega _{BD})(4+3\omega _{BD})}\frac \alpha
{\alpha_0}\propto t^n,  \label{phi1}
\end{equation}
so $\alpha G$ is always a constant. Note that for large values of
$\omega _{BD}$ we have a simple relation between the values of $G$
and $\alpha$:

\begin{equation}
G\frac \alpha {\alpha_0} \approx \frac{-\zeta \omega _{BD}}{\pi
\omega }>0 \label{key}
\end{equation}
As expected, $\alpha $ increases whilst $G$ falls as $t\rightarrow
\infty $ in a flat universe. It is interesting to note that the
asymptotic value of $G\alpha$ is uniquely determined by the
parameters in the model with no arbitrary constants.

\begin{figure}[htb]
\psfig{file=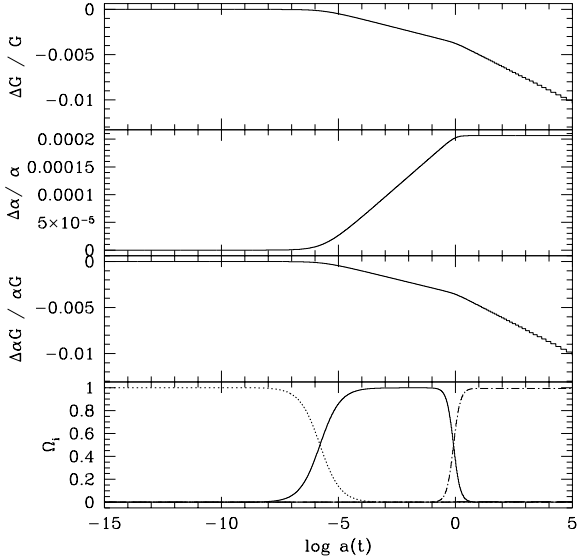,height=8cm} \caption{Evolution of the
relative shift in the values of the two 'constants' in a realistic
flat cosmology with vacuum energy and approximately accurate
initial values for the fields. We start from a radiation-dominated
universe where both $\alpha$ and $G$ stay constant. Thereafter we
move into dust domination where $\alpha$ changes slowly, while $G$
goes through a small decrease. As the universe becomes dominated
by the vacuum energy, $\alpha$ goes to a constant, while $G$ goes
on decreasing indefinitely as in ordinary Brans Dicke theory.
Values used for the couplings are the minimum allowed value of
$\omega_{BD} = 3500$ and we take the best fit value of $\zeta /
\omega = -10^{-4}$ from BSBM theory.} \label{ldom}
\end{figure}

\begin{figure}
\psfig{file=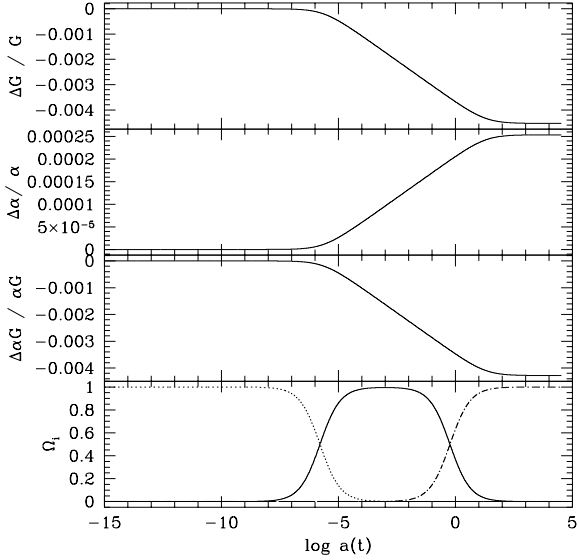,height=8cm} \caption{Evolution of the
relative shift in the values of the two 'constants' in an open
universe through radiation, dust and curvature-dominated epochs.
Initial values for the fields are set so as to give realistic
values at present time. Again $\alpha$ and $G$ are constants in
the radiation dominated era, whilst $\alpha$ increases and $G$
decreases through matter domination. As curvature starts to take
over the expansion, both $\alpha$ and $G$ tends to constants.
Values for the couplings are the minimum allowed value of
$\omega_{BD} = 3500$ and the best fit value of $\zeta / \omega =
-10^{-4}$ from the BSBM theory.} \label{kdom}
\end{figure}

Although the asymptotic behaviour is now determined, the question
of whether this can be reached on a cosmological timescale depends
strongly on the choice of initial conditions and needs to be
investigated numerically. We can quickly conclude that the
asymptotic regime is not reached in our universe. Presently we
have $\alpha \approx 1/137$, and in our units the numerical value
of $M$ is extremely small, $\sim 10^{-123}$.  Obviously the actual
value of $\alpha$ is then many orders of magnitude larger than
given by the solution in eq.(\ref{psi1}) and we are thus nowhere
near the asymptotic regime. Consequently, in order to find the
behaviour of $\alpha$ and $G$ we turn to numerical solutions of
the equations. We evolve the Friedmann equations through time with
initial conditions chosen so as to yield the present day values of
$G$ and $\alpha$. For $\alpha$ we find a behaviour very similar to
the BSBM theory, with a slow growth giving a relative change of
the order $10^{-4}$ throughout the dust epoch. $G$ goes through a
decrease of order $10^{-3}$ during the same period. The numerical
results are shown in Figures \ref{ldom} and \ref{kdom}.


\section{Radiation era evolution}

The evolution in the radiation era is slightly more complicated because of
the contribution of the $\rho _\psi $ term to the right-hand side of the $%
\phi $ evolution equation. This means that we do not have the usual
late-time asymptotic behaviour of constant $\phi $ to accompany the $%
a=t^{1/2}$ scale factor as in BD radiation universes. If we assume

\[
a=t^{1/2}
\]
then we have

\begin{equation}
\frac d{dt}(\dot \phi t^{3/2})=R-\lambda t^{3/2}\dot \psi ^2  \label{newphi}
\end{equation}
where

\begin{eqnarray}
\lambda &\equiv &\frac{8\pi \omega }{3+2\omega _{BD}},  \label{lam} \\
R &\equiv &\frac{8\pi M}{3+2\omega _{BD}}=-\frac{4\pi \omega N}{\zeta
(3+2\omega _{BD})}  \label{R}
\end{eqnarray}

The $R$ term is negligible when the kinetic energy of the $\psi $
field dominates the matter density during the radiation era.
Likewise, the $\lambda$ term can be neglected when the matter
density dominates the $\psi $ kinetic energy. We also have

\begin{equation}
\frac d{dt}(\dot{\psi}t^{3/2})=N\exp [-2\psi ]  \label{p}
\end{equation}
as in the case with constant $G.$ This has the exact solution

\begin{equation}
\psi =\frac 12\ln (8N)+\frac 14\ln (t)  \label{psi3}
\end{equation}
as before, so $\dot \psi ^2=(16t^2)^{-1}$. If we substitute this in (\ref
{newphi})

\begin{equation}
\frac d{dt}(\dot{\phi}t^{3/2})=R-\frac \lambda {16}t^{-1/2}  \label{q}
\end{equation}
so

\begin{equation}
\phi =\phi _0+2Rt^{1/2}-\frac \lambda 8\ln (t)+Ct^{-1/2}  \label{phi3}
\end{equation}
where $C$ and $\phi _0$ are constants. If the universe expands for long
enough to reach the asymptotic regime then we have (as $R>0)$

\begin{eqnarray}
\psi &\approx &\frac 14\ln (t)  \label{s1} \\
\phi &=&G^{-1}\approx -\frac{8\pi \omega N}{\zeta (3+2\omega _{BD})}t^{1/2}
\label{s2}
\end{eqnarray}
so, from eqns. (\ref{key}) and (\ref{alf}),

\begin{equation}
G\frac \alpha {\alpha_0} \approx \frac{-\zeta (3+2\omega
_{BD})}{8\pi \omega N}\approx \frac{ -\zeta \omega _{BD}}{4\pi
\omega N}  \label{key2}
\end{equation}
for large $\omega _{BD}$. Thus we still have the nice asymptotic
behaviour of $\alpha G$ in the radiation-dominated epoch. However,
we again need to compare with numerical results to determine
whether these asymptotic solutions can indeed be realised in the
Universe. As in the case of dust, the same simple reality check
can now be performed on the solution (\ref{psi3}). As in the case
of constant $G$ we are nowhere near this particular solution in
our Universe. $\alpha$ would need to be several orders of
magnitude smaller if it was to satisfy the solution, and as in the
BSBM theory we expect instead a constant value of $\alpha$ in the
rad epoch. This assumption is indeed confirmed by the numerical
solutions shown in Figures (\ref{kdom}) and (\ref{ldom}).

Another possible problem for the analytic solutions above would
arise if the kinetic energy of the $\psi$ field dominates the
matter density during radiation domination. We regard this
situation as unrealistic and it cannot be realised asymptotically.


\section{Curvature era evolution}

During a curvature-dominated phase of an open universe the expansion scale
factor tends to that of the Milne vacuum universe, which is an exact
solution of general relativity and of Brans-Dicke theory (with constant $%
\phi $) with

\begin{equation}
a=t  \label{mil}
\end{equation}

Using this in the propagation equations for $\phi $ and $\psi ,$we find
that, as $t\rightarrow \infty $, so leading order

\begin{eqnarray*}
\psi &=&\psi _{*}-\frac{N\exp [-2\psi _0]}t \\
\phi &=&\phi _{*}+\frac{4\pi \omega N}{\zeta (3+2\omega _{BD})t} \\
&&
\end{eqnarray*}
with $\psi _{*}$ and $\phi _{*}$ constants, so both $\alpha $ and
$G$ tend to constant values as $t\rightarrow \infty $. In a
universe that passes directly from dust domination to curvature
domination these constant values will be very close to the
asymptotic attractors for the dust era of evolution found above in
eqn. (\ref{key}) providing the dust epoch has lasted long enough
for the attractor to be reached.

The behaviour of $G, \alpha$ and $G \alpha$ in a universe like our
own but which eventually becomes dominated by negative curvature
is shown in Figure (\ref{kdom}).

\section{Cosmological 'constant' era evolution}\label{lambda}

In flat Brans-Dicke cosmologies a solution of the Friedmann equation with
cosmic vacuum energy ($p_v=-\rho _v)$ is

\begin{eqnarray}
a &=&t^{\omega _{BD}+\frac 12}  \label{v1} \\
\phi &=&\phi _ot^2  \label{v2} \\
\phi _0 &\equiv &\frac{32\pi \rho _v}{(5+6\omega _{BD})(3+2\omega _{BD})}
\label{v3}
\end{eqnarray}
and $\rho _v$ is constant. This is not the general solution but it is the
attractor for the general $p_v=-\rho _v$ solution at late times \cite{bm}.
It is a power-law inflation model \cite{stein}. Note that in Brans-Dicke
theory, unlike in general relativity, a $p_v=-\rho _v$ stress behaves
differently in the Friedmann equation to an explicit constant $\Lambda $ term%
\cite{bm}. It is the former that describes the stress contributed by a
stationary scalar field with a constant potential. Every term in the BD
Friedmann equations falls as $t^{-2}$ for this solution. It is unusual in
that it appears to predict that if the universe has just begun accelerating
(as observations imply, \cite{super,efst}) then $G$ should vary rapidly in
the solar system. However, this argument assumes that the vacuum stress is
dominant everywhere, right down to the solar system scale, which in reality
it is not.

If we substitute this solution for $a(t)$ (but not $\phi $) in the $\psi $
and $\phi $ evolution equations, (\ref{phi}) and (\ref{ddotpsi}) then we
get, since

\[
p_v=-\rho _v=const,
\]
that

\begin{eqnarray*}
\ddot \phi +\frac{3(2\omega _{BD}+1\ )}{2t}\dot \phi &=&\frac{8\pi }{%
3+2\omega _{BD}}(4\rho _v-2\rho _\psi )\approx \frac{-8\pi }{3+2\omega _{BD}}%
(4\rho _v-\omega \dot \psi ^2) \\
\ddot \psi +\frac{3(2\omega _{BD}+1\ )}{2t}\dot \psi &=&-\frac 2\omega
e^{-2\psi }\zeta \rho _m\approx 0
\end{eqnarray*}
So, at late times

\begin{eqnarray*}
\dot \psi &=&Aa^{-3}=Dt^{-3(\omega _{BD}+\frac 12)} \\
\psi &=&Et^{-3\omega _{BD}-\frac 12}+F\rightarrow F \\
&& \\
\ddot \phi +\frac{3(2\omega _{BD}+1\ )\ }{2t}\dot \phi &=&\frac{-8\pi \omega
\ }{3+2\omega _{BD}}(4\rho _v\ -Qa^{-6})\rightarrow \frac{-32\pi \omega \rho
_v\ }{3+2\omega _{BD}}, \\
&&\  \\
&&
\end{eqnarray*}
so

\begin{equation}
\phi =A+Bt^2+Ct^{-3\omega _{BD}-\frac 12}\rightarrow \phi _0t^2,
\label{phi4}
\end{equation}

and, as expected, we get the same growing behaviour as in pure BD.
When the universe becomes vacuum-energy dominated $\alpha $ tends
to a constant value but $\alpha G\propto G\propto t^{-2}$
continues to fall. This behaviour is confirmed by numerical
solutions shown in Figure (\ref{ldom}). Using eqn. (\ref {key}),
we see that if $t_{v\text{ }}$is the time when a vacuum-dominated
era succeeds a sufficiently long dust-dominated era in a flat
universe, then at $t\geq t_v$ in the vacuum-dominated era we
expect

\begin{equation}
\alpha (t)G(t)=\frac{-\zeta \omega _{BD}}{\pi \omega }\left( \frac{t_v}t%
\right) ^2.  \label{ag1}
\end{equation}
Hence, today, we would have

\begin{equation}
\alpha (t_0)G(t_0)=\frac{-\zeta \omega _{BD}}{\pi \omega (1+z_v)^{\frac 4{%
1+2\omega _{BD}}}}.\ \   \label{ag2}
\end{equation}
We see that, as in the situation where $G$ is constant, the effect
of a vacuum energy or quintessence field is to turn off variations
in $\alpha $ when it takes over the expansion of the universe,
\cite{sbm}.

\section{Discussion}

We have formulated a simple gravity theory which extends general relativity,
by the addition of two scalar fields, to include time variation of $G$ and $%
\alpha$. Previously, the study of the cosmological variation of
physical 'constants' has confined attention to varying one
constant only or to discussing the effects of altering the values
of physical constants without a self-consistent theory for their
dynamical variation \cite{land}, \cite {moffat93}. The structure
of unified gauge theories and particle physics theories with extra
dimensions has given some indication as to the self consistency
conditions required if traditional constants are allowed to vary,
\cite{marc,drink}.

We have found that the expansion of the universe is affected by
varying $G$ to first order and the evolution of the expansion
scale factor follows the behaviour found in Brans-Dicke
cosmologies to leading order without being significantly affected
by variations in $\alpha$. The variations in $\alpha $ are
affected by the variations in $G$ through their influence on the
expansion rate. This is significant in the dust-dominated era of
cosmic expansion, which is known to exhibit a special mathematical
behaviour in the absence of $G$ variation, \cite{bsm}. The effect
of any $G$ time variation simplifies the $\alpha $ variation and
allows an exact solution to be found with $\alpha \propto
t^n,$where $n\equiv 2/(4+3\omega _{BD})\ $is determined by the
Brans-Dicke parameter $\omega _{BD}$.



In both the radiation and dust dominated eras, there are
asymptotic solutions in which the product $G\alpha $ remains
constant and its value is determined uniquely by the coupling
constants of the theory. However in universes like our own with
the values of $\alpha$ and $G$ near present values these
asymptotic regimes are not reached throughout the life of the
universe. Typically our present values for $\alpha$ are much
larger than the values required by the asymptotic solution.

In a curvature-dominated or quintessence-dominated era the
variation in $\alpha $ ceases, just as in the situation with no
$G$ variation \cite{sbm}. This is an important feature of all
models with varying $\alpha $ in theories of the BSBM sort because
it naturally reconciles evidence of variations in $\alpha $ at
redshifts $z\sim 1-3$, with local ($z=0.1)$ constraints from the
Oklo natural reactor if the universal expansion began to
accelerate at $z\sim 0.7,$ as current observations imply.

Finally, we reiterate that the conclusions drawn above apply only
to varying-$\alpha $ theories with negative $\zeta$. The exact
solutions given in eqs. (\ref{psi1}),(\ref{alf2}),and (\ref{psi3})
for the evolution of $\alpha (t)$ during the radiation and dust
eras no longer exist when $\zeta >0$ and hence $N<0.$ During the
curvature and cosmological constant-dominated eras the evolution
of $a(t)$ is significantly changed by the variations of $\psi $
and the assumptions (\ref{mil}) and (\ref{v1}) for the scale
factor evolution are no longer valid.

The study performed here provides a simple cosmological model in
which the variation of two 'constants' can be studied exactly. A
number of extensions are possible. The variations of weak and
strong couplings can be included and the constraints imposed by
any scheme grand unification can be imposed \cite{marc}.

{\bf Acknowledgements} We thank Bernard Carr and David Mota for
discussions. HBS acknowledges support from the Research Council of
Norway. The numerical work shown herein was performed on COSMOS,
the Origin 2000 supercomputer owned by the UK-CCC.

\end{document}